\documentclass[12pt]{article}
  \usepackage{amsfonts}
  \usepackage{amsmath}
\usepackage{amssymb}
\usepackage{amscd}
\usepackage{graphicx}
\begin{document}

~~
\bigskip
\bigskip
\begin{center}
{\Large {\bf{{{Pauli energy spectrum for twist-deformed space-time}}}}}
\end{center}
\bigskip
\bigskip
\bigskip
\begin{center}
{{\large ${\rm {Marcin\;Daszkiewicz}}$}}
\end{center}
\bigskip
\begin{center}
\bigskip

{ ${\rm{Institute\; of\; Theoretical\; Physics}}$}

{ ${\rm{ University\; of\; Wroclaw\; pl.\; Maxa\; Borna\; 9,\;
50-206\; Wroclaw,\; Poland}}$}

{ ${\rm{ e-mail:\; marcin@ift.uni.wroc.pl}}$}

\end{center}
\bigskip
\bigskip
\bigskip
\bigskip
\bigskip
\bigskip
\bigskip
\bigskip
\bigskip
\begin{abstract}
In this article, we define the Pauli Hamiltonian function for twist-deformed N-enlarged
Newton-Hooke space-time provided in article \cite{nnh}. Further, we derive its energy spectrum, i.e., we find the
corresponding eigenvalues as well as the proper eigenfunctions.
\end{abstract}
\bigskip
\bigskip
\bigskip
\bigskip
\eject

The suggestion to use noncommutative coordinates goes back to
Heisenberg and was firstly  formalized by Snyder in \cite{snyder}.
Recently, there were also found formal  arguments based mainly  on
Quantum Gravity \cite{2}, \cite{2a} and String Theory models
\cite{recent}, \cite{string1}, indicating that space-time at the Planck
scale  should be noncommutative, i.e., it should  have a quantum
nature. Consequently, there appeared a lot of papers dealing with
noncommutative classical and quantum  mechanics (see e.g.
\cite{mech}, \cite{qm}) as well as with field theoretical models
(see e.g. \cite{prefield}, \cite{field}), in which  the quantum
space-time is employed.

In accordance with the Hopf-algebraic classification of all
deformations of relativistic \cite{class1} and nonrelativistic
\cite{class2} symmetries, one can distinguish three basic types
of space-time noncommutativity (see also \cite{nnh} for details):\\
\\
{ \bf 1)} Canonical ($\theta^{\mu\nu}$-deformed) type of quantum space \cite{oeckl}-\cite{dasz1}
\begin{equation}
[\;{ x}_{\mu},{ x}_{\nu}\;] = i\theta_{\mu\nu}\;, \label{noncomm}
\end{equation}
\\
{ \bf 2)} Lie-algebraic modification of classical space-time \cite{dasz1}-\cite{lie1}
\begin{equation}
[\;{ x}_{\mu},{ x}_{\nu}\;] = i\theta_{\mu\nu}^{\rho}{ x}_{\rho}\;,
\label{noncomm1}
\end{equation}
and\\
\\
{ \bf 3)} Quadratic deformation of Minkowski and Galilei  spaces \cite{dasz1}, \cite{lie1}-\cite{paolo}
\begin{equation}
[\;{ x}_{\mu},{ x}_{\nu}\;] = i\theta_{\mu\nu}^{\rho\tau}{
x}_{\rho}{ x}_{\tau}\;, \label{noncomm2}
\end{equation}
with coefficients $\theta_{\mu\nu}$, $\theta_{\mu\nu}^{\rho}$ and  $\theta_{\mu\nu}^{\rho\tau}$ being constants.\\
\\
Moreover, it has been demonstrated in \cite{nnh}, that in the case of the
so-called N-enlarged Newton-Hooke Hopf algebras
$\,{\mathcal U}^{(N)}_0({ NH}_{\pm})$ the twist deformation
provides the new  space-time noncommutativity of the
form\footnote{$x_0 = ct$.},\footnote{ The discussed space-times have been  defined as the quantum
representation spaces, so-called Hopf modules (see e.g. \cite{oeckl}, \cite{chi}), for the quantum N-enlarged
Newton-Hooke Hopf algebras.}
\begin{equation}
{ \bf 4)}\;\;\;\;\;\;\;\;\;[\;t,{ x}_{i}\;] = 0\;\;\;,\;\;\; [\;{ x}_{i},{ x}_{j}\;] = 
if_{\pm}\left(\frac{t}{\tau}\right)\theta_{ij}(x)
\;, \label{nhspace}
\end{equation}
with time-dependent  functions
$$f_+\left(\frac{t}{\tau}\right) =
f\left(\sinh\left(\frac{t}{\tau}\right),\cosh\left(\frac{t}{\tau}\right)\right)\;\;\;,\;\;\;
f_-\left(\frac{t}{\tau}\right) =
f\left(\sin\left(\frac{t}{\tau}\right),\cos\left(\frac{t}{\tau}\right)\right)\;,$$
$\theta_{ij}(x) \sim \theta_{ij} = {\rm const}$ or
$\theta_{ij}(x) \sim \theta_{ij}^{k}x_k$ and  $\tau$ denoting the time scale parameter
 -  the cosmological constant. Besides, it should be  noted, that the  above mentioned quantum spaces {\bf 1)}, { \bf 2)} and { \bf 3)}
 can be obtained  by the proper contraction limit  of the commutation relations { \bf 4)}\footnote{Such a result indicates that the twisted N-enlarged Newton-Hooke Hopf algebra plays a role of the most general type of quantum group deformation at nonrelativistic level.}.

In this article we investigate the impact of the twisted N-enlarged
Newton-Hooke space-time \cite{nnh}\footnote{In the formula (\ref{relations}) $\kappa_a$
denotes the deformation parameter such that $\lim_{\kappa_a \to 0}f_{\kappa_a}(t)=0$.}
\begin{eqnarray}
[\;\hat{ x}_{1},\hat{ x}_{2}\;] = if_{\kappa_a}({t})\;,\label{relations}
\end{eqnarray}
on the Pauli energy spectrum \cite{pauli}\footnote{It means, that from physical point of view, we investigate the impact of high-energy (transplanckian) regime
on the one of the most important quantum system, such as the nonrelativistic electron with spin $s=\frac{1}{2}$, moving in constant
electric and magnetic fields simultaneously.}. 
Firstly, however, we remaind the basic facts
concerning the Pauli model defined on classical space. In this aim we start with the following canonical commutation relations for momentum and position
operators $(x_i,p_i)$
\begin{equation}
[\;x_i,x_j\;] = 0 =[\;p_i,p_j\;]\;\;\;,\;\;\; [\;x_i,p_j\;]
={i\hbar}\delta_{ij}\;, \label{classcom}
\end{equation}
Then, the Hamiltonian (Pauli) function for nonrelativistic electron with spin, moving in
the external electric field ${\bar E} = -{\rm grad} \phi$ and in the
magnetic field ${\bar B} = {\rm rot} {\bar A}$,
is defined as a sum of two mutually commuting terms
\begin{equation}
H=H_1(\bar{p}, \bar{x})+H_2(\bar{x},\bar{\sigma})\;, \label{ham}
\end{equation}
such that\footnote{The symbols $m$, $e$, $c$ denote mass, electric charge and speed of light respectively.}
\begin{equation}
H_1(\bar{p}, \bar{x}) = {1\over 2m}\left({\bar p}+{e\over c}{\bar A}(\bar{x})\right)^2
-e\phi(\bar{x})\;, \label{ham1}
\end{equation}
and
\begin{equation}
H_2(\bar{x},\bar{\sigma}) = \frac{1}{2m}\bar{\sigma}\cdot \bar{B}(\bar{x})\;, \label{ham2}
\end{equation}
with Pauli matrices vector $\bar{\sigma} = (\sigma_1,\sigma_2,\sigma_3)$. Besides, it is easy to see, that in so-called symmetric gauge framework
\begin{eqnarray}
{\bar A}(\bar{x})=\left[-{B\over 2}x_2,{B\over 2}x_1,0\right]\;, \label{os2}
\end{eqnarray}
and for the following choice of the electric field $\phi(\bar{x})$
\begin{eqnarray}
\phi(\bar{x}) =-Ex_1\;, \label{os3}
\end{eqnarray}
the above operators take the form\footnote{From now, for simplicity, we consider electron moving in $(x_1,x_2)$-plane.}
\begin{eqnarray}
H_1(\bar{p}, \bar{x}) &=&
\frac{1}{2m}\left[
\left[ p_1 -\frac{eB}{2c} x_2 \right]^2 +
\left[ p_2 +\frac{eB}{2c} x_1 \right]^2 \right] +eEx_1\;, \label{ham1a}\\
H_2(\bar{x},\bar{\sigma}) &=& H_2(\sigma_3) \;=\; \frac{1}{2m}{\sigma}_3B = \frac{B}{4m}
\begin{bmatrix}
  1 & 0   \\
  0 &-1
\end{bmatrix}
\;. \label{ham2a}
\end{eqnarray}
The energy spectrum of first term (\ref{ham1a}) has been found with use of creation/annihilation operator technique in paper
\cite{energylandau}. It looks as follows
\begin{eqnarray}
{H}_1(\bar{p}, \bar{x}) \psi_{(n,\alpha)}(\bar{x}) = {{E}}_{(n,\alpha)}\psi_{(n,\alpha)}(\bar{x})\;. \label{eigen1}
\end{eqnarray}
where
\begin{eqnarray}
\psi_{(n,\alpha)}(\bar{x}) &=& \exp{i\left[\alpha x_2+{m\omega\over 2\hbar}x_1x_2\right]}\cdot \frac{1}{\sqrt{(2m\hbar\omega)^n n!}}(a^{\dag})^n|0>\;,\label{states}\\
{{E}}_{(n,\alpha)} &=& {\hbar\omega\over 2}(2n+1)-\frac{1}{m}\left[
{\hbar\lambda}\alpha-{\lambda^2\over 2}\right]\;,\label{energies}
\end{eqnarray}
with $n=0,1,2, \ldots$, real parameter $\alpha$, cyclotron frequency $\omega=\frac{eB}{mc}$, parameter $\lambda=\frac{mcE}{B}$ and with
$(a,a^{\dag})$-objects given by
\begin{eqnarray}
&~~&a^\dag =-2i{p}^*+{eB\over 2c}x+\lambda \;\;\;,\;\;\;
a =2i{p}+{eB\over 2c}x^*+\lambda\;, \nonumber\\
&~~&x={x}_1+i{x}_2\;\;\;,\;\;\; {p} =\frac{1}{2} ({p}_1
-i{p}_2)\;, \label{oper1a}\\
&~~&(\alpha + i\beta)^* = \alpha -i\beta \;\;\;,\;\;\; a|0> =0\;.\nonumber
\end{eqnarray}
The eigenvectors and eigenvalues of the second term (\ref{ham2a}) can be find as well. They take the form
\begin{eqnarray}
{H}_2(\sigma_3) \psi_{\pm}(\bar{x}) = {{E}}_{\pm}\psi_{\pm}(\bar{x})\;. \label{eigen2}
\end{eqnarray}
where
\begin{eqnarray}
\psi_+ =
 \begin{pmatrix}
  1  \\
  0
 \end{pmatrix}
\;\;\;,\;\;\;
\psi_- =
 \begin{pmatrix}
  0  \\
  1
 \end{pmatrix}
\;\;\;,\;\;\;
{{E}}_{\pm} = \pm \frac{B}{2m} \label{spinorial2}
\end{eqnarray}
Consequently, the energy spectrum of the whole Hamiltonian operator (\ref{ham}) looks as follows
\begin{eqnarray}
H(\bar{p}, \bar{x}) = \psi_{\pm(n,\alpha)}(\bar{x}) = {{E}}_{\pm(n,\alpha)}\psi_{\pm(n,\alpha)}(\bar{x})
\;, \label{wholeham1a}
\end{eqnarray}
with
\begin{eqnarray}
\psi_{\pm(n,\alpha)}(\bar{x}) = \psi_{(n,\alpha)}(\bar{x})\otimes \psi_{\pm}
\;, \label{wholestate}
\end{eqnarray}
and
\begin{eqnarray}
{{E}}_{\pm(n,\alpha)} = {{E}}_{(n,\alpha)} + {{E}}_{\pm} = {\hbar\omega\over 2}(2n+1)-\frac{1}{m}\left[
{\hbar\lambda}\alpha-{\lambda^2\over 2} \pm \frac{B}{2}\right]
\;, \label{wholeenergy}
\end{eqnarray}
respectively.

Let us now turn to the main aim of our investigations - to the derivation of Pauli energy levels for quantum space-times (\ref{relations}).
For this purpose, we extend the deformed spaces to the whole algebra of momentum and position operators as follows
\begin{eqnarray}
&&[\;\hat{ x}_{1},\hat{ x}_{2}\;] = if_{\kappa_a}({t})\;\;\;,\;\;\;
[\;\hat{ p}_{i},\hat{ p}_{j}\;] =0\;\;\;,\;\;\;[\;\hat{ x}_{i},\hat{ p}_{j}\;] = {i\hbar}\delta_{ij}\;. \label{phasespaces1}
\end{eqnarray}
Next, by analogy to the commutative case, we define the twist-deformed Hamiltonian operator as (see (\ref{ham1a}), (\ref{ham2a}))
\begin{eqnarray}
\hat{H}(t) &=& \hat{H}(\bar{\hat{p}},\bar{\hat{x}}) = \frac{1}{2m}\left[
\left[ \hat{p}_1 -\frac{e}{2c}{\check{B}(f_{\kappa_a}({t}))} \hat{x}_2 \right]^2 +
\left[ \hat{p}_2 +\frac{e}{2c}{\check{B}(f_{\kappa_a}({t}))} \hat{x}_1 \right]^2 \right]\;+\label{grom1}\\
&+&e\hat{E}(f_{\kappa_a}({t}))\hat{x}_1 +
\frac{1}{2m}\sigma_3\hat{B}(f_{\kappa_a}({t}))
\;,\nonumber
\end{eqnarray}
where
\begin{eqnarray}
\check{B}(f_{\kappa_a}({t})) &=& \frac{2}{f_{\kappa_a}(t)}\left[(1-f_{\kappa_a}(t)B)^{-\frac{1}{2}}-1\right]\;,\\
\hat{B}(f_{\kappa_a}({t})) &=& \frac{B}{1+f_{\kappa_a}({t})B}\;,\\
\hat{E}(f_{\kappa_a}({t})) &=& \frac{E}{1+f_{\kappa_a}({t})E}\;.
\end{eqnarray}
It should be noted, that the form of the above magnetic inductions $(\check{B}(f_{\kappa_a}({t})),\hat{B}(f_{\kappa_a}({t})))$ as well as the electric field $\hat{E}(f_{\kappa_a}({t}))$
is dictated by the results of papers \cite{string1} and \cite{ncfields}, which concern the classical electrodynamics defined for the canonical noncommutativity  (\ref{noncomm}).
Then, as we shall see for a moment, for the most simple case $f_{\kappa_a}({t})=\kappa_a=\theta={\rm const.}$, we reproduce from the obtained in present article energy levels the correct $\theta$-deformed Pauli spectrum provided in \cite{ncfields}. Besides, one can check, that in terms of commutative variables $(x_i,p_i)$ the operator (\ref{grom1}) takes the form\footnote{We use
the standard link between both types of operators defined by: $\hat{x}_i=x_i-\epsilon_{ij}
\frac{f_{\kappa_a}({t})}{2\hbar}p_j$ and $\hat{p}_i=p_i$ (see e.g. \cite{link}); the calculations are performed in the same way as in article \cite{daszlandau}.}
\begin{eqnarray}
\hat{H}(t) = \hat{H}_1(t) + \hat{H}_2(t)
\;,\label{grom2}
\end{eqnarray}
with
\begin{eqnarray}
\hat{H}_1(t) &=&
\frac{1}{2m}\left[
\left[ (1-\alpha_{\kappa_a}(t) ){p}_1 -\frac{e}{2c}{\check{B}(f_{\kappa_a}({t}))} {x_2} \right]^2 \right.\;+\nonumber\\
&~~&\;\;\;\;\;\;\;\;\;\;\;\;+\;\left.\left[ (1-\alpha_{\kappa_a}(t) ){p}_2 +\frac{e}{2c}{\check{B}(f_{\kappa_a}({t}))} {x_1} \right]^2 \right] + \label{grom2} \\
&+&e\hat{E}(f_{\kappa_a}({t}))\left[{x_1}-{f_{\kappa_a}(t)\over 2\hbar}{p}_2\right] = \hat{H}(\bar{{p}},\bar{{x}},t)
\;,\nonumber\\
\hat{H}_2(t) &=& \frac{1}{4m}{\sigma}_3\hat{B}(f_{\kappa_a}({t})) = \hat{H}_2(\sigma_3,t)\;\;\;,\;\;\;
\alpha_{\kappa_a}(t) = \frac{e}{4\hbar c}{f_{\kappa_a}(t) \check{B}(f_{\kappa_a}({t}))}\;,
\end{eqnarray}
and
\begin{eqnarray}
[\;\hat{H}_1(t),\hat{H}_2(t)\;] = 0
\;.\label{grom3}
\end{eqnarray}
The spectrum of the above $\hat{H}_1$-term can be derived with use of ladder operator scheme in the same manner as in paper \cite{daszlandau}. It is given by
\begin{eqnarray}
\hat{H}_1(t)\psi_{(n,\alpha,\kappa_a)}={{E}}_{(n,\alpha,\kappa_a)}(t)\psi_{(n,\alpha,\kappa_a)}(t)
\;, \label{grom33}
\end{eqnarray}
where $n=0,1,2, \ldots$, $\alpha \in \mathbf{R}$ and
\begin{eqnarray}
\psi_{(n,\alpha,\kappa_a)}(t) &=&
\exp{i\left[\alpha x_2+{m{\omega(t)}\over 2\hbar\beta^2(t)}x_1x_2\right]}\cdot{({(2m\hbar{\omega}(t))^n n!})}^{-1/2}\;\cdot
\nonumber \\
&\cdot&({a}^{\dag}(t))^n|0>\;,\nonumber\\
{{E}}_{(n,\alpha,\kappa_a)}(t)&=&
{\hbar{\omega(t)}\over 2}(2n+1)-
\frac{1}{m}{\hbar\beta(t)\lambda_+(t)}\alpha-{m\over 2}\lambda_-^2(t)
\;, \nonumber\\
a^\dag(t) &=&-2i{\bar{p}}^*(t)+\frac{e}{2c}{\check{B}(f_{\kappa_a}({t}))}x+\lambda_{-}(t)\;, \label{grom9h}\\
a(t) &=& 2i{\bar{p}}(t)+\frac{e}{2c}{\check{B}(f_{\kappa_a}({t}))}x^*+\lambda_{-}(t)\;,\nonumber\\
\beta(t)  &=& (1 - \alpha_{\kappa_a}(t))\;\;\;,\;\;\;\lambda_{\pm}(t)= \lambda \pm
\frac{em}{4\beta(t) \hbar}{\hat{E}(f_{\kappa_a}({t}))f_{\kappa}(t)}\;,\nonumber\\
\omega(t) &=& \beta(t)\omega\;\;\;,\;\;\;{\bar{p}}(t)=\beta(t)p(t)
\nonumber\;.
\end{eqnarray}
The eigenvalues and eigenvectors of the second term $\hat{H}_2(t)$ are defined by the equation (\ref{spinorial2}).
Consequently, due to the commutation relations (\ref{grom3}), the twist-deformed (Pauli) energy levels look as follows
\begin{eqnarray}
\psi_{\pm(n,\alpha,\kappa_a)}(t) &=&\exp{i\left[\alpha x_2+{m{\omega(t)}\over 2\hbar\beta^2(t)}x_1x_2\right]}\cdot
{({(2m\hbar{\omega}(t))^n n!})}^{-1/2}\;\cdot \nonumber\\
&\cdot&({a}^{\dag}(t))^n|0>\otimes \;\psi_{\pm}\;, \label{final1} \\
{{E}}_{\pm(n,\alpha,\kappa_a)}(t)&=&{{E}}_{(n,\alpha,\kappa_a)}(t) + E_{\pm} =
{\hbar{\omega(t)}\over 2}(2n+1)-
\frac{\hbar}{m}{\beta(t)\lambda_+(t)}\alpha\;+\nonumber\\
&-&{m\over 2}\lambda_-^2(t) \pm \frac{1}{2m}{\hat{B}(f_{\kappa_a}({t}))}
\;.\label{final2}
\end{eqnarray}
It is easy to see, that in obvious way they depend on time, and such a property probably follows from nonstationary character of the background space-time noncommutativity (\ref{relations}). In other words, from physical point of view, the above mentioned attribute can be interpreted as an direct impact of quantum space dynamics just on the total energy of moving particle. However, unfortunately, the deeper understanding of such a behavior of the discussed system seems to be at this stage rather difficult, and it is postponed for the future investigations. Apart of that, one should also observe, that the spectrum (\ref{final1}) and (\ref{final2}) structurally appears very similar to its commutative counterpart, given by the  formulas (\ref{wholestate}) and (\ref{wholeenergy}), i.e., it still remains linear for example in frequency parameter $\omega(t)$, and it stays quadratic in function $\lambda_-(t)$. Probably such a feature follows from the specific, central charge-like form of the commutation relations (\ref{relations}), and in the case of another, nontrivially deformed quantum spaces, it should be rather lost. Of course, for $f_{\kappa_a}(t) = 0$ the above energy levels become classical, while for the most simple (canonical) space-time noncommutativity (\ref{noncomm}) - given by $f_{\kappa_a}({t})=\theta$ - they reproduce for $E=0$ the results of article \cite{ncfields}.

\section*{Acknowledgments}
The author would like to thank J. Lukierski for valuable discussions.
 This paper has been financially  supported  by Polish
NCN grant No 2014/13/B/ST2/04043.

\end{document}